# Oil economy phase plot: a physical analogy

Luciano Celi[b], Claudio Della Volpe[a], Luca Pardi[b], Stefano Siboni[a]

[a] Department of Civil, Environmental and Mechanical Engineering, University of Trento – Italy
[b] Institute for Chemical and Physical Processes, National Research Council, Pisa – Italy

**Abstract**

A phase plot of the oil economy is built using the literature data of world oil production, price, and EROEI (Energy Returned on Energy Invested). An analogy between the oil economy and the Bénard convection is proposed; some methods of interpretation and forecast of the system behaviour are also shown based on "phase portrait" using as main variables the price, production and EROEI values. The scenery is proposed on this basis.

## 1. Introduction

The term phase plot or phase portrait has two main meanings. In physical chemistry, it is used to indicate the traditional equilibrium and transitions among the different phases of a system, while in mechanics, ecology, economy, irreversible thermodynamics, or chemical kinetics it is used to show the pathways of a dynamical system in the space of its natural variables. In the present work, the term will be used in this second and wider meaning.

A dynamical system is represented by a set of differential equations, whose solution is often only numerically assessable; the solutions may be more easily represented in a phase space, whose dimension is given by the number of the natural variables of the system. From this point of view, extremely different dynamical systems may show similar behaviours and therefore this approach may be helpful to characterize complex and not well-understood systems. For our aims, it is worthwhile to consider the behaviours of systems represented in 2D and 3D phase portraits, sometimes considered as projections of more complex, higher-dimensional systems. The interested reader can find many detailed analyses available in the literature [1–3].

Here we summarize some important points.

The time is not explicitly represented in the plot but it is possible to follow the time evolution of the system variables corresponding to different conditions; the time sequence generally shows a trajectory around some well-defined points or sets, called attractors; the number of trajectories is potentially infinite or very large; the plot may be usually divided into various regions corresponding to different combinations of variables and directions of the movement; the attractors may correspond to stable or unstable subsets of states (each state corresponding to points of the phase space).

The procedure is easily applied in mechanical systems such as the pendulum [3], but it can be similarly applied in ecology (prey-predator cycles, the Lotka-Volterra equations [10]) or chemical processes such as oscillating reactions (the Belousov-Zhabotinsky reaction [4, 5]) or economic analysis (as the Goodwin representation of the economic cycle [6]).

Even when the system appears clearly more complex and a 2D representation is only a 2D projection of a dimensionally more complex plot, it is possible to have very useful indications even in absence of a well-known set of equations that govern the system dynamics.

This is the case of the Etkin representation of the Earth climate [11] (fig. 1), where the two main variables, whose correlation is extremely complex, are the $CO_2$ concentration and the Earth mean temperature as extracted by different data sources. The plot shows that in recent years the Earth system has abandoned a strong

"attractor" active for at least one million years to explore a different space phase territory. It is apparent, even from an easy consideration of the Etkin image that no a priori full knowledge of the system is necessary to appreciate the worth and importance of the process.

Some other systems appear to be irreducibly 3D (or even more dimensionally complex) as it is the case of the Lorentz "strange attractor" and other Lorenz-like systems, which will be considered in the following.

In the present analysis, we try to use a procedure similar to that suggested by Etkin's work, although concerning a completely different topic: the oil production system. We will try to show that 2D and 3D phase portraits may be useful to shed light on the dynamic's subtleties of such a system.

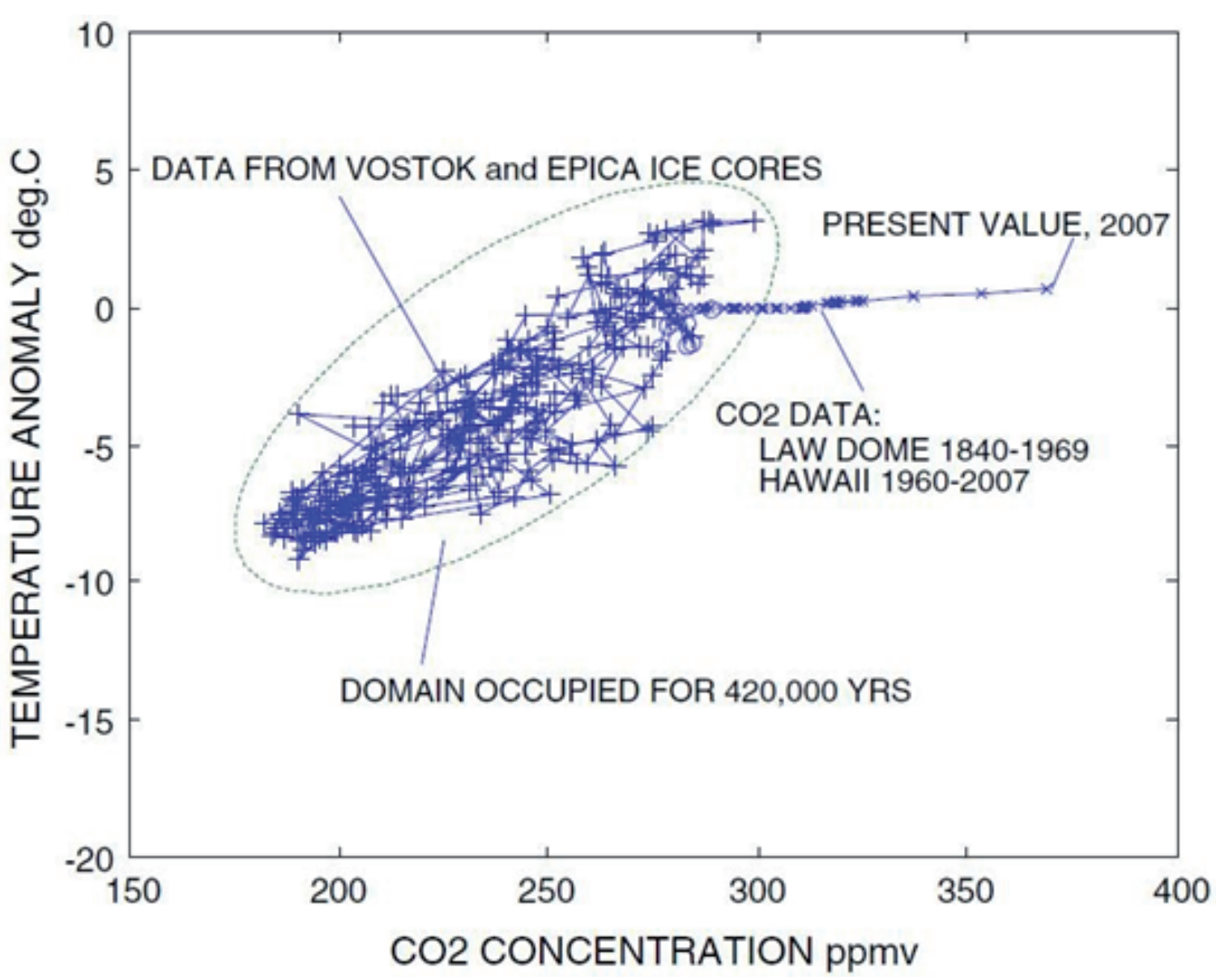

*Fig. 1: State-space view of Antarctic ice-age cycles, from Etkin, 2010 (License Number: 3971331483376 by Springer Editor)*

It is also clear that, if we assume that the system is described by a dynamical model and therefore its parameters and equations do not depend explicitly on time, the curve overlapping is impossible unless the phase plot is a 2D projection of a system with a higher dimension. Nevertheless, significant conclusions may be derived also from this low dimensional plot.

## 2. Data and methods

Following the Etkin approach, we try here to build a phase plot based on the significant variables of the oil production system hoping to have some insight into the behaviour of the system. Such a system is certainly very complex and the basic problem is to choose, if they exist, few variables able to capture the essential behaviour of it. This choice may depend on theoretical and empirical considerations; for the data considered here, we have chosen the price of oil (constant price 2014 $) and the annual global oil production in Mton (millions of tons). This combination of data is available in the literature of the oil economy and particularly in one of the most popular public databases regularly updated by BP [12]. Such a choice of the variables is not casual, however. Indeed, it is fully justified by the possible analogy between thermodynamics and economy proposed by Saslow [7]. To represent the economic wealth, Saslow, trying to explore the analogies between the Economy and Thermodynamics, has proposed the following equation for the wealth:

$$W=LM+pN \text{ [eq. 1]}$$

where the wealth W is a first-order homogeneous function of the money amount M and goods quantity

N; the other two quantities are respectively L, the value of money and p, the value of goods. In other terms, the wealth is available partly as goods of different kinds and partly also as money, whose value changes in time. The value of money or that of goods is the derivative of the wealth with respect to the quantity of money or goods, respectively.

In this way, p and L play the role of intensive quantities, while M and N are extensive quantities. These functions are very similar and have the same mathematical homogeneity properties of the thermodynamic functions such as energy, entropy, Helmholtz or Gibbs free energies, as analysed in the common approach introduced more than 50 years ago by Callen and used today in most thermodynamics textbooks [13].

In the specific treatment proposed by Saslow, the number of goods and its derivative p are equivalent to the number of moles and to the chemical potential, respectively, while W may be considered as an analogue of the Helmholtz free energy. From this point of view, a plot of the price vs. the quantity of a good, will correspond to the plot of the chemical potential of a reactant versus the number of moles obtained from a chemical reaction.

## 3. Results

In fig. 2 it is reported the yearly mean oil price (at constant 2014 $) versus the annual oil production in Mton - in the model each parameter could be made adimensional dividing it by its value in a reference year. It is quite obvious that the quantities are in fact "continuous" from a mathematical point of view but that they are yearly sampled. It is possible to find in literature similar plots with similar data, but none has been considered from the present point of view [8,14], as a phase plot of a dynamical system. In [8], however, the use of monthly data for a very short time makes the plot very noisy and prevents understanding the global trend, while [14] data (starting from the half of the XIX century until the present) are not commented.

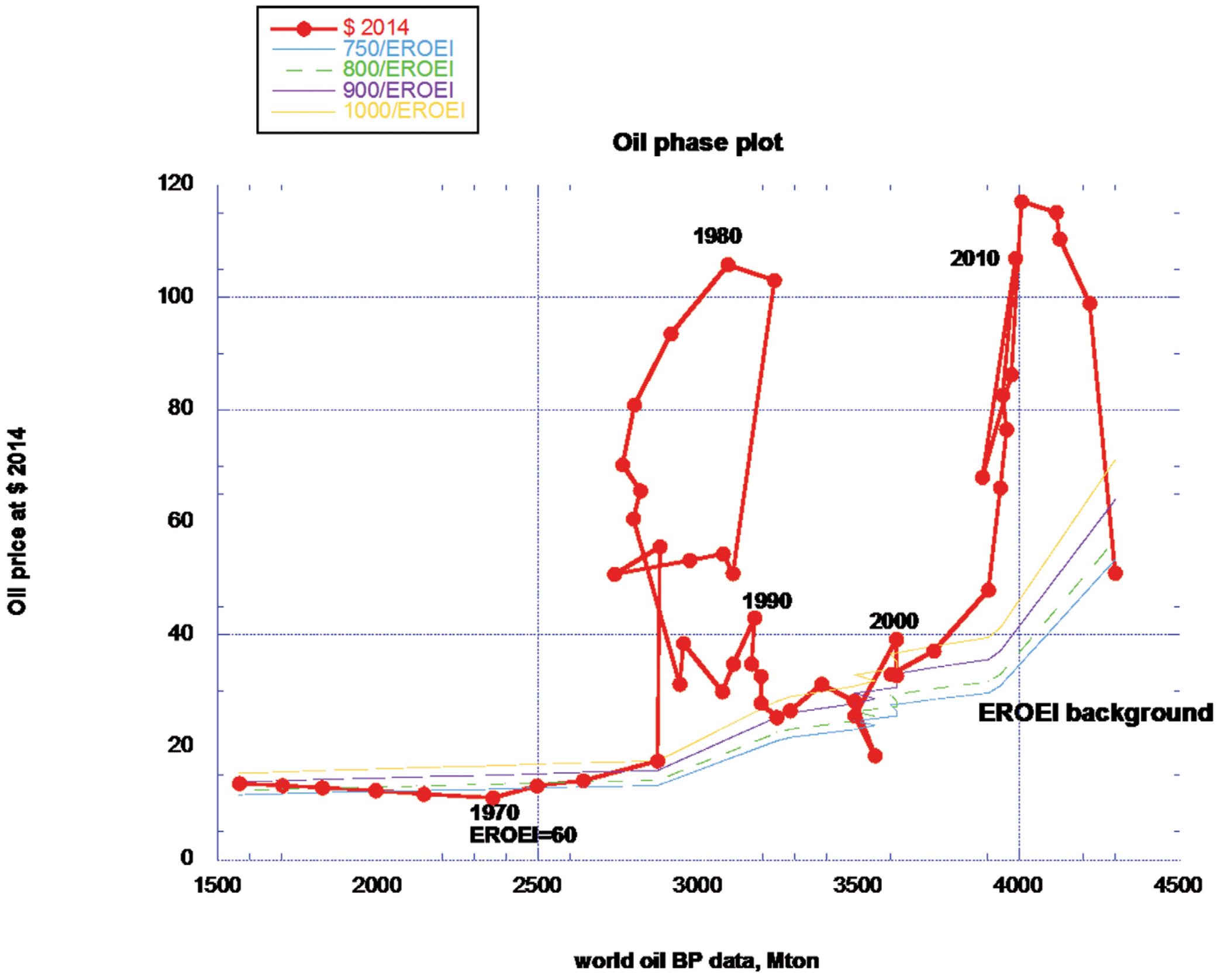

*Fig. 2: 2D phase plot, oil production (Mton) vs oil price ($ 2014). The background curves in different colors correspond to different limit prices at EROEI=1.*

In the present plot, the yearly mean values are connected by a red timeline with a well-defined pathway, but the time is not explicitly present as a variable. The fact that the curve intersects itself indicates that at least a third variable is necessary. Some general comments on the plot can be made:

1. The plot has been obtained using a very limited number of points with respect to the very high and potentially infinite number of a classical phase plot; this is a matter of fact and cannot be changed. It probably constitutes the main limitation in applying the method to a typical economic system; the number of points can be increased using monthly or even daily values, but this introduces an unacceptable noise in the trajectory.

2. The p vs N trajectory seems to rest on a growing background identified by the lowest points of the plots, grossly indicated by the first six points and by four colored curves. This increasing "background" could be "fitted" using a function of the form: P=k/(EROEI), where EROEI is the Energy Return on Energy Invested as estimated in literature [11], the functional form corresponds to the assumption that, barring any perturbation of the oil market, the price of oil is determined by its energy cost up to a constant factor. An acceptable agreement was obtained with reasonable k values; the parameter k, whose units are $/barrel could be considered the limit price of this economic sector. If EROEI=1 which is the theoretical limit for the use of oil as the energy source the hypothetical price is k. The effect of EROEI is to reduce this limit price to the effective market price. Noticeably this fitting corresponds to assume that EROEI and price are, at least approximately, inversely proportional to the background curve, i.e. that their product is a constant. Such a remark will be useful in the following. The introduction of this parameter has been done on a purely empirical basis and starting from the knowledge of the long term evolution of this and other strategic goods of mineral origin and substantially non-renewable resources.

3. The two "anomalies" above the fitted background are the most striking features of the plot and correspond to the two major historical crises of the oil market and world economy. In the first one, the yearly production reduces for a certain time interval, while in the second case this has been not true. In a plot as that of ref. 3 this kind of anomaly is simply described as a non-elastic behaviour of the oil market, losing the details of the exact trajectory and the movement direction of the system. After each one of those anomalies, the trajectory returns along the background line. The background curve describes the pathway of the system in the absence of specific events as the crises; the zone below the background curve is forbidden by technical reasons, while the zone above the background is normally forbidden by market reasons; during crises events, the system goes above the background curve. The interpretation of the dynamic trajectory could be that at some critical values of the oil production the system is not able immediately to increase the production and follow the market request, yielding an abrupt increase of the price at a constant production level. Such an occurrence may induce the search for an oil source able to enter the market at the new price level; if the search is successful the oil production may even reduce (because one can obtain a new source, thanks to the technological advancement, even at a higher EROEI; this is probably the cause of the oil production reduction during the first crisis).

The lowest price was obtained in the same year (1998) in which some authors calculated the highest EROEI at a world level in that time interval [15]. However, the existence of some crossing points implicitly forces one to introduce a third variable, at least, and EROEI seems to be a reasonable choice, even if its equivalent thermodynamic meaning is not clear at all at this point of the analysis. A possible 3D plot is illustrated below, by assuming a linear reduction of EROEI in the considered time interval (Fig. 3).

In this new representation of the same data, the different behavior of the two crises may be described as a rotation direction. The presence of such a distinct behavior, opposite rotation directions in the same phase plot, is not a common feature. It looks very similar to what is found in the 2D projection of a very special 3D dynamical system with three variables, the well-known Lorenz attractor [9]. This apparently minor detail has its origin in the deep structure of the Lorenz plot; in the original system, which has been found useful in very different fields, the modeled phenomenon is the so-called Bénard convection, possibly occurring in a thin layer of a liquid heated from below in a gravitational field. When heating transport overcomes certain critical values, the simple thermal diffusion is replaced by a complex and structured pattern of convective cells, whose size and characteristics depend on the physical parameters of the system.

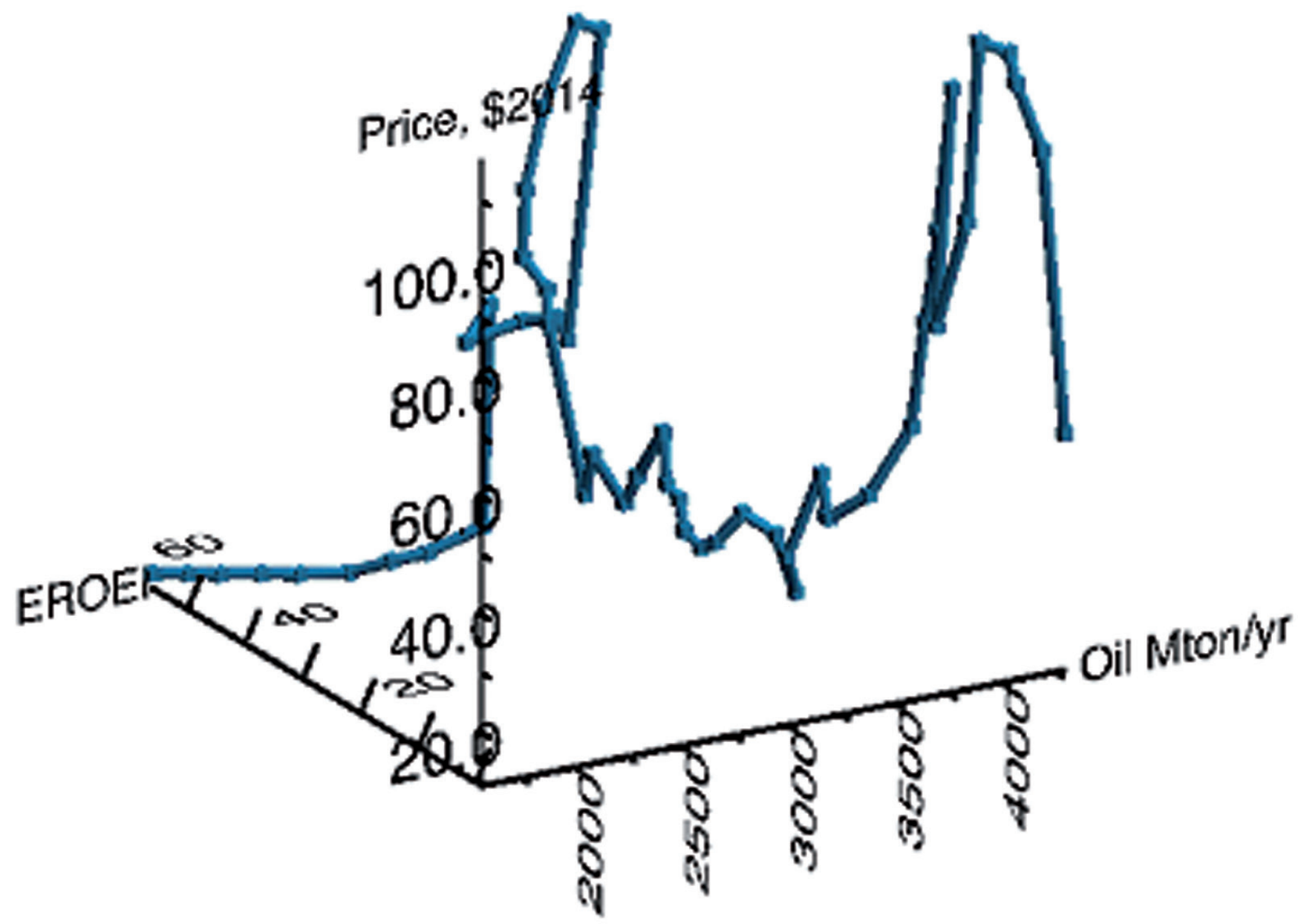

*Fig. 3: 3D phase plot (looking from the X-axis, Oil production). The rotation directions are opposite. The EROEI is supposed to linearly decrease from the beginning; oil production (Mton); oil price ($ 2014).*

This consideration may be the starting point to mark the analogy between these two systems. There is an idea already discussed in the literature [16] according to which the industrial revolution, using fossil fuels, enormously increased the energy flow through human society and the oil market is the most recent phase of this increased flow. In a similar way as the convection greatly enhances the thermal flow compared to heat diffusion giving rise to the Bénard convection, this quantitative increase in the energy flow through human socioeconomic system gives rise to emerging phenomena, described by the present plot.

Due to the roughly similar, general behavior of Lorenz's and oil systems, we can try to further outline the analogy by correlating the three Lorenz parameters to the three parameters of the present plot.

| Dimension | Lorenz paper | This paper |
|---|---|---|
| X | Convective intensity. | Oil production, N. |
| Y | $\Delta T$ between ascending and descending currents, similar signs of X and Y denoting that warm fluid is riding and cold fluid descending. | EROEI, E. |
| Z | Proportional to the deviation of the vertical temperature profile from linearity, a positive value indicating that the largest gradients occur near the boundaries. | Oil price, P. |

The analogy between convective intensity and oil production comes from the idea that energy flow in human society changed from a slow diffusion process to a faster convective movement; the analogy between price and the non-linear behaviour of the temperature comes from the similar behaviour of the two variables: in fact "the strongest gradients occur near the boundaries" (original Lorenz observation) [9] also for the price values, as suggested by the [14] data, where the largest price variations appear at the beginning of oil market (about 1850) and during the last two crises.

There is another consideration coming from the fitting of the "background" curve. In that fitting the product of price and EROEI is approximately constant; is the equivalent product of Y*Z in Lorenz approach approximately constant too? The answer is positive; in fact, while Z has its maxima at the boundaries, the Y parameter is maximum at the centre of the considered interval.

Such qualitative analogy between the two systems may be explored quantitatively, searching for appropriate estimates to the parameters of the Lorenz equations as applied to the oil system. It should be considered that the number of points is limited, only 50 points, or a few more than 100 using the data of ref. 14. In both cases however the final result is not satisfactory.

Indeed, it is anyway possible to evaluate the three parameters of the Lorenz model, even if it is unreasonable to hope for a good correspondence. In fact, the three parameters, evaluated with very few figures of merit of the fit, would be:

| Qualitative dimension | Lorenz paper | This paper |
|---|---|---|
| Prandtl number | Pr= 28 | K1= −0.009±0.008 |
| Relative Rayleigh number | r= 10 | K2=1.47±0.07 |
| Geometric parameter | b= 8/3 | K3=0.37±0.08 |

These values can provide no information about any possible “strange” or chaotic behaviour of the system, but only about a possible common single attractor. Moreover, it is not possible to find any simple meaning for the evaluated parameter.

As a conclusion, the analogy between oil production and the Lorenz system appears to remain only qualitative, perhaps for lack of data, while the possibility of an attractor may be considered. The existence of such an attractor may be confirmed by considering another phase plot built dividing the data of oil production by the world population [17]; the price versus mean per capita production plot is given in Fig. 4.

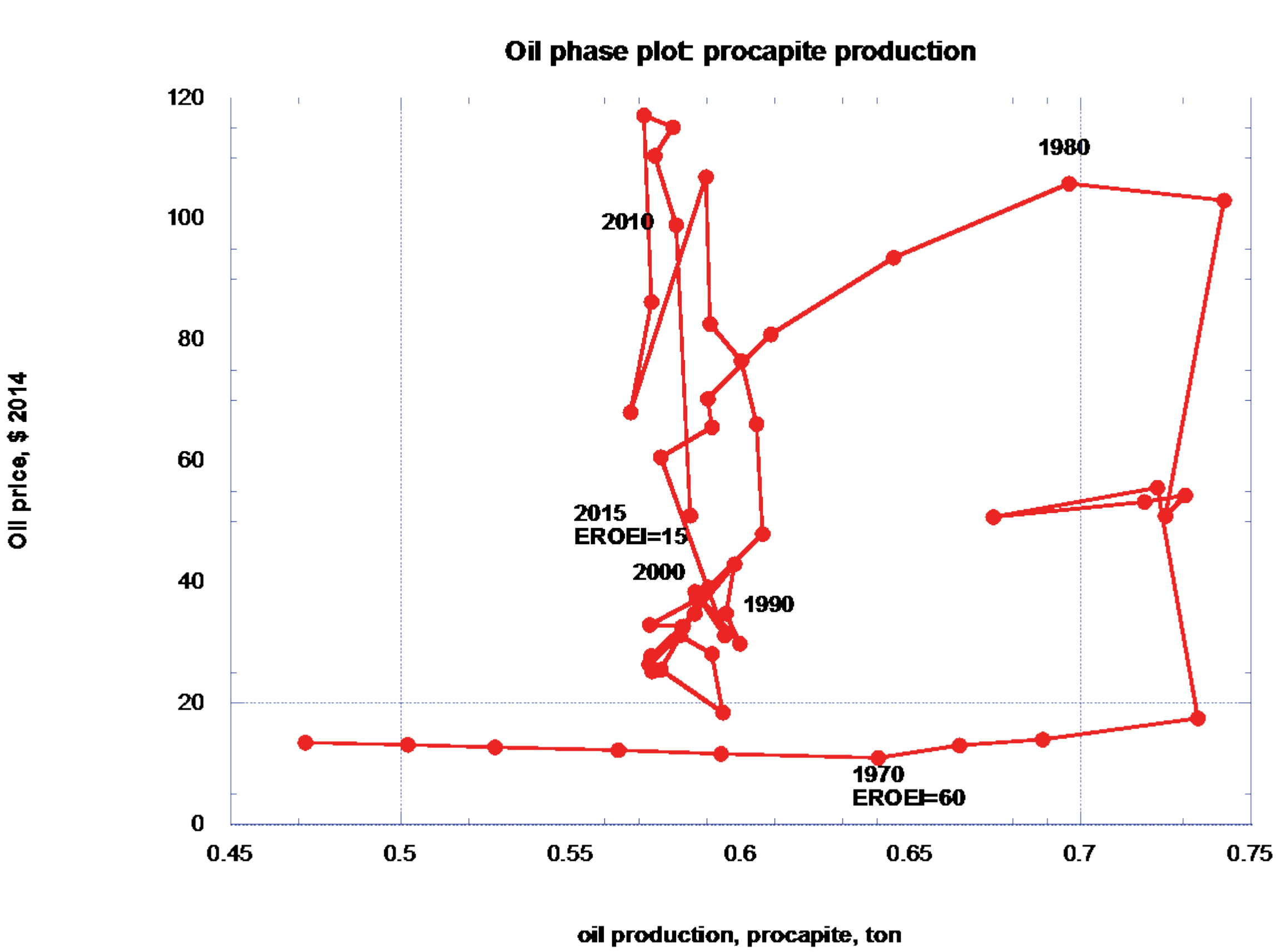

*Fig. 4: Phase plot of procapite oil production (Mton) vs oil price ($ 2014).*

From this plot it is apparent that the last 50 years of oil production are "attracted" toward a value of about 0.59 ton per capita; a value which seems to oscillate only from 0.57 to 0.61 in the last 25 years; a possible interpretation may be that the oil use in human society is very "rigid", or that in other terms it is impossible to reduce this value unless profound modifications of the technology in critical sectors (sea transports, agriculture production, mobility) are attained.

This rigidity in the use of oil could help to understand the behavior of the background curve; the function k/(EROEI) using the analyzed data may help to foresee the future of oil market and production. The final consideration may concern the possible prediction of the oil market scenery based on the present analogical model:

a) the EROEI may only decrease (in the mean and on the intermediate time-interval of 1-5 years) and accordingly, the price may only increase save for instantaneous oscillations due to external components of the market, and for a strong reduction of consumption due to substitution and/or economic recession;

b) when the fitting will diverge to infinity (vertical curve) or better when the corresponding "basic" price will approach values above 100$ (2014), the situation may become very critical.

Accepting this analogy, one may estimate some parameters for this apparently inescapable evolution of the oil market; using the fitting curve $P_{background}$= k/Eroei (k may vary between 750 and 100) one may evaluate that an EROEI of 4-5 may correspond to a "background" price of 100 $ 2014; this point could be approached for a yearly production of 5000-5500 Gton, with a prediction of 15-25 years from today. This point should not simply mark an "oil peak" but a real full depletion crisis; obviously this foresee is based on a simple analogy and we don't claim this scenery is nothing more than an analogy.

But even this simple analogy forces us to consider that the only strategy, which is also in agreement with the climatic situation so well described recently in Paris (COP 21) [18], is to move with a strong effort toward a society based on renewable energies, mainly in the critical sectors indicated above.

***Acknowledgements***

We are grateful to dr. Laherrere for some of the original numeric data considered in this paper.